\documentclass{PoS}

\title{Dynamical Local Chirality and Chiral Symmetry Breaking}

\ShortTitle{Dynamical Local Chirality and Chiral Symmetry Breaking}

\author{Andrei Alexandru\\
        George Washington University, Washington, DC, USA\\
        E-mail: \email{aalexan@gwu.edu}}

\author{\speaker{Ivan Horv\'ath}\\
        University of Kentucky, Lexington, KY, USA\\
        E-mail: \email{horvath@pa.uky.edu}}

\abstract{We present some of the reasoning and results substantiating 
the notion that spontaneous chiral symmetry breaking (SChSB) in QCD is encoded 
in local chiral properties of Dirac eigenmodes. Such association is possible 
when viewing chirality as a {\em dynamical effect}, measured with respect 
to the benchmark of statistically independent left--right components. 
Following this rationale leads to describing local chiral behavior by 
a taylor--made {\em correlation}, namely the recently introduced correlation 
coefficient of polarization $C_A$. In this language, 
correlated modes ($C_A>0$) show dynamical preference for local chirality while 
anti--correlated modes ($C_A<0$) favor anti--chirality. Our conclusion is that
SChSB in QCD can be viewed as dominance of low--energy correlation (chirality) 
over anti--correlation (anti--chirality) of Dirac sea. The spectral range of 
local chirality, chiral polarization scale $\Lambda_{ch}$, is a dynamically 
generated scale in the theory associated with SChSB. One implication of these 
findings is briefly discussed.}

\FullConference{Xth Quark Confinement and the Hadron Spectrum\\
		8-12 October 2012\\
		TUM Campus Garching, Munich, Germany}

\begin{document}

\section{Introduction}

The aim of this work is to gain insight into the phenomenon of spontaneous 
chiral symmetry breaking in QCD.~\footnote{Since this presentation at 
{\em Confinement X}, a more comprehensive account of this work has been given 
in Ref.~\cite{Ale12D}.}
Despite being one of the cornerstones in hadronic physics, standing virtually 
uncontested, SChSB remains shrouded in clouds with regard to its dynamical nature. 
The quark dynamics of QCD is represented by Dirac eigenmodes and at least 
the low--energy part of this spectrum is accessible via numerical lattice QCD. 
It is thus both interesting and practical to study the features of low--lying 
Dirac eigensystem in some detail. For example,
it could prove fruitful to inquire what distinguishes low--lying eigenmodes 
in broken theory from those in symmetric one.

A well-known spectral distinction is that the eigenmodes of a broken theory condense, 
while those of a symmetric theory do not. Indeed, invoking its spectral representation, 
scalar fermionic density in the massless limit (``chiral condensate'') is proportional 
to the spectral density of near--zeromodes (``mode condensate''), the fact known as 
the Banks--Casher relation~\cite{Ban80A}. However, being an implicit definition 
of SChSB, mode condensation feature is more of a kinematical constraint than 
a window into the dynamical specifics of the breaking mechanism.

In going beyond the Banks--Casher relation, it is natural to examine the spinorial 
structure of the eigenmodes and, in particular, to look for the imprints of broken 
chiral symmetry in their {\em chiral properties}. The roots of our present approach 
go back to Ref.~\cite{Hor01A} in that the intention is to characterize the behavior 
of the eigenvectors {\em locally}. Indeed, global chiral properties of Dirac modes 
are fixed and with acceptable lattice discretization, such as overlap 
fermions~\cite{Neu98BA} used here, this is faithfully reproduced at 
the regularized level. The local behavior, on the other hand, reflects details 
of dynamics induced by interacting quarks and gluons.

Nevertheless, a significant conceptual change had to take place~\cite{Ale10A} for 
local chirality to become a useful tool in bottom--up approach to QCD vacuum 
structure~\cite{Hor06A}, i.e. in meaningful characterization of QCD vacuum properties 
without reference to models. This change has to do with viewing chirality 
as a dynamical concept, quantified relative to the situation when left and right 
components are independent degrees of freedom. A framework for characterizing 
dynamical properties of generic polarization phenomena has been built around this 
idea~\cite{Ale10A}, with absolute $X$--distribution $P_A(X)$ and 
correlation coefficient of polarization $C_A$ being the associated quantifiers.

Our main point here is that {\em dynamical chirality} of near--zeromodes, quantified
by $C_A$, may provide for a dynamical spectral signature faithfully distinguishing 
chirally broken situation from the symmetric one~\cite{Ale12D}. Specifically, in 
the broken case, a band of locally polarized modes ($C_A>0$) occupies the spectral 
region around the surface of the Dirac sea that is otherwise anti--chiral ($C_A<0$). 
In transition to the symmetric case, the polarized band dissolves into the sea and 
there is only anti--chirality. Picturing this in reverse, and using properly defined 
observables, SChSB in QCD--like theories takes on the meaning of 
{\em condensing chirality}~\cite{Ale12D}.  

There are two points we wish to highlight with regard to this scenario. First,
it involves a natural scale, namely the width of a polarized band: 
the {\em chiral polarization scale} $\Lambda_{ch}$~\cite{Ale10A}. In view of the above, 
$\Lambda_{ch}$ can in fact be viewed as an ``order parameter'' of SChSB, and its 
existence has several interesting consequences~\cite{Ale12D}. Secondly, like mode 
condensation, chirality condensation is not necessarily tied to the masslessness of 
dynamical quarks. Indeed, mode condensation is equivalent to existence of {\em valence} 
chiral condensate irrespective of dynamical quark masses. The meaning of our proposal 
is similar in that dynamical chirality condenses and $\Lambda_{ch}$ is generated if and 
only if valence condensate is non--zero, and thus whenever modes themselves condense.

In this presentation, we discuss the concept of dynamical chirality and focus on some of 
the basic building blocks of the above picture, namely on establishing that $\Lambda_{ch}$ 
is a true dynamical scale and that the situation in N$_f$=2+1 QCD is consistent with it.  
 
\section{Dynamical Polarization and Dynamical Local Chirality}

We first summarize some needed basic elements of the dynamical polarization 
framework~\cite{Ale10A}. Assume that a probabilistic object takes values in the linear space 
that can be decomposed into a pair of equivalent orthogonal subspaces, so that its
``sample'' $Q$ can be written as $Q=Q_1+Q_2$ with $Q_1 \cdot Q_2=0$. Dynamics of 
this object is encoded in its probability distribution function ${\cal P}_f(Q_1,Q_2)$, 
and our goal is to adjudicate whether it supports outcomes favoring asymmetric 
participation of the two subspaces (polarization) or the symmetric participation 
(anti--polarization). Note that we are not interested in the overall preference of one 
subspace over the other (global polarization), which is non--existent for cases studied 
here that satisfy ${\cal P}_f(Q_1,Q_2)={\cal P}_f(Q_2,Q_1)$. Rather, we are interested in 
characterizing ``sample polarization'', quantifying asymmetry without distinguishing 
which subspace happens to prevail in any given $Q$. When samples are labeled by position 
coordinates, the term {\em local polarization} becomes appropriate.

\begin{figure} 
\centerline{
\includegraphics[width=.5\textwidth]{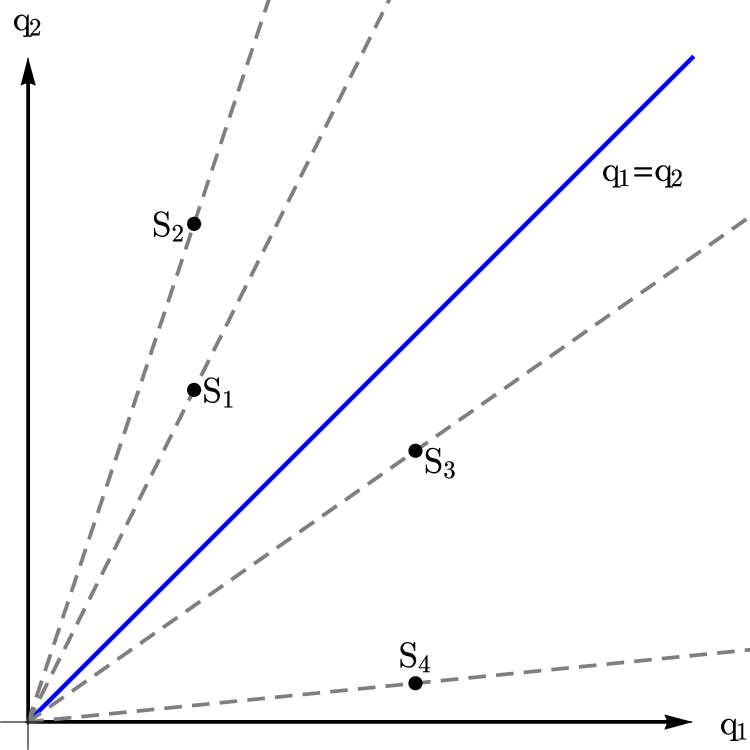}
} 
\caption{Rays of equal polarization in the sample space of ${\cal P}_b(q_1,q_2)$.
The solid blue line ($q_2=q_1$) consists of points with zero polarization. Relative
polarization of various samples is a uniquely defined concept.} 
\label{fig:1} 
\end{figure} 

Participation of a given subspace in sample $(Q_1,Q_2)$ is measured by the magnitude
of its component $q_i \equiv |Q_i|$, and it is thus sufficient to consider the associated
distribution of magnitudes ${\cal P}_f(Q_1,Q_2) \rightarrow {\cal P}_b(q_1,q_2)$. 
The sets of equally polarized points in this restricted sample space are the rays 
$q_2 = t q_1$, as shown in Fig.~\ref{fig:1}. Since the subspaces are equivalent, there 
is a well--defined ray $q_2=q_1$ of unpolarized samples, and the degree of polarization 
has to grow symmetrically away from this reference. However, other than that, 
the assignment of polarization value to any given ray -- the polarization function -- is 
not fixed a priori, and polarization characteristics of dynamics ${\cal P}_f(Q_1,Q_2)$ 
will in general depend on it. 

While it is not possible to uniquely assign the value of polarization to samples,
such as those shown in Fig.~\ref{fig:1}, we can say with certainty e.g. that $S_2$
is more polarized than $S_1$, that $S_1$ is more polarized than $S_3$, or that $S_2$ 
is less polarized than $S_4$. This has more than trivial meaning when samples to be 
compared come from two different dynamics, namely two different distributions 
${\cal P}_b(q_1,q_2)$. In that case this is telling us that relative polarization 
characteristics based on sample--wise comparisons are absolute: they are invariant 
under the choice of polarization function~\cite{Ale10A}.

Our goal is to define a {\em dynamical} polarization measure where ``dynamical'' means
{\em correlational} and invariant in the above sense. The essence of correlation is that
it is referenced to statistical independence, which leads us to sample--wise comparisons
with the associated distribution ${\cal P}_b^u(q_1,q_2)$ of independent components. More 
precisely, if $p(q)\equiv \int d q' {\cal P}_b(q,q') = \int d q' {\cal P}_b(q',q)$ is 
the marginal distribution of a component, then ${\cal P}_b^u(q_1,q_2)=p(q_1) p(q_2)$.
The simplest dynamical characteristic of polarization can then be constructed as follows.
Imagine simultaneous drawings of samples from ${\cal P}_b(q_1,q_2)$ and 
${\cal P}_b^u(q_1,q_2)$ and keeping score of their polarization comparisons. 
The result of such experiment is the probability $\Gamma_A$ that a sample produced
by dynamics under consideration is more polarized than sample from uncorrelated 
distribution. The correlation coefficient of polarization is then
\begin{equation}
  C_A \equiv 2 \Gamma_A -1    \qquad C_A \in [-1,1]
  \label{eq:10}
\end{equation}
Thus, polarization--enhancing dynamics ($\Gamma_A>1/2$) are positively correlated
while polarization--suppressing dynamics ($\Gamma_A<1/2$) are anti--correlated.
One can also define a more detailed dynamical polarization measure, namely absolute 
$X$--distribution $P_A(X)$, based on differential comparisons to statistical 
independence~\cite{Ale10A}.

The above can be straightforwardly applied to study of dynamical local chirality 
in Dirac eigenmodes. Indeed, the two subspaces in question are the left and right 
spinorial subspace and so $Q=Q_1+Q_2 \rightarrow \psi(x)=\psi_L(x) + \psi_R(x)$. 
The collection of local values $\psi(x)$ in a given mode or a group of modes provide
samples representing the distribution ${\cal P}_f(Q_1,Q_2)$. To characterize 
the Dirac spectrum in terms of dynamical local chirality, we define the average 
correlation of polarization at scale $\lambda$ in finite volume $V$ as
\begin{equation}
   C_A(\lambda,M,V) \equiv 
   \frac{\sum\limits_k \langle \, \delta(\lambda - \lambda_k) \, C_{A,k} \,\rangle_{M,V}}
        {\sum\limits_k \langle \, \delta(\lambda - \lambda_k) \, \rangle_{M,V}}
   \,=\,
   \frac{\rho_{ch}(\lambda,M,V)}{\rho(\lambda,M,V)}
    \label{eq:20}
\end{equation}
where $C_{A,k}$ is the correlation of $k$-th mode, $M$ labels dynamical quark masses, and  
\begin{equation}
   \rho_{ch}(\lambda,M,V) \equiv \frac{1}{V} \, 
   \sum\limits_k \langle \, \delta(\lambda - \lambda_k) \, C_{A,k} \,\rangle_{M,V}
   \label{eq:30}
\end{equation} 
is the spectral polarization density defined in analogy to usual spectral density 
$\rho(\lambda)$ of modes~\cite{Ale12D}. In this formal language, the theory exhibits
mode condensation if $\lim_{\lambda \to 0} \lim_{V\to \infty} \rho(\lambda,M,V) > 0$,
and it exhibits {\em dynamical chirality condensation} if 
$\lim_{\lambda \to 0} \lim_{V\to \infty} \rho_{ch}(\lambda,M,V) > 0$. Note that 
the negativity of the latter would imply condensation of anti--chirality.

\begin{figure}[t]
\begin{center}
    \centerline{
    \hskip 0.12in
    \includegraphics[width=0.52\textwidth,angle=0]{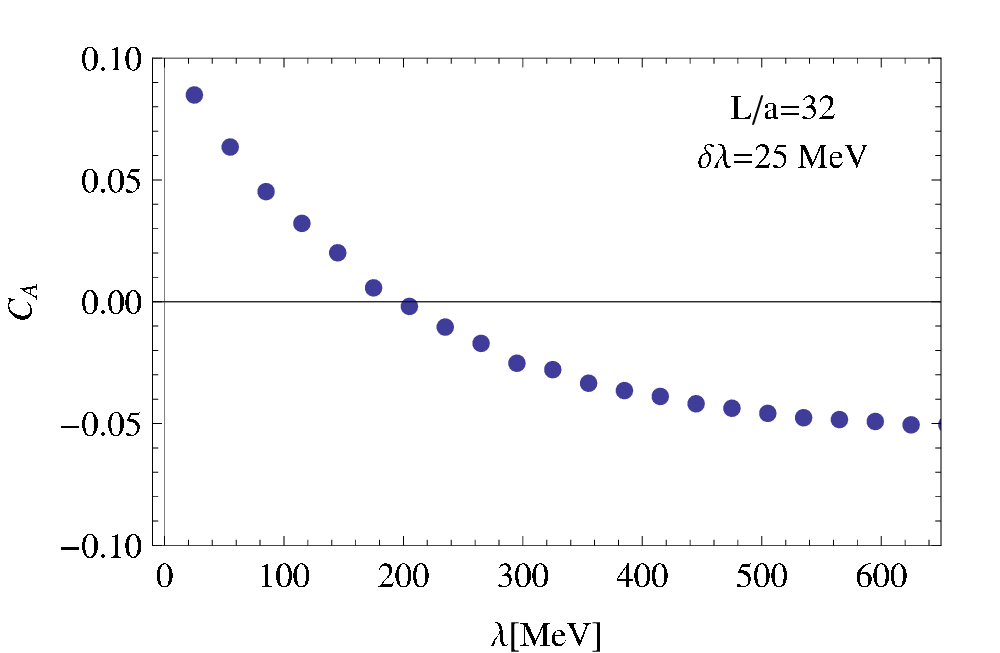}
    \hskip -0.10in
    \includegraphics[width=0.52\textwidth,angle=0]{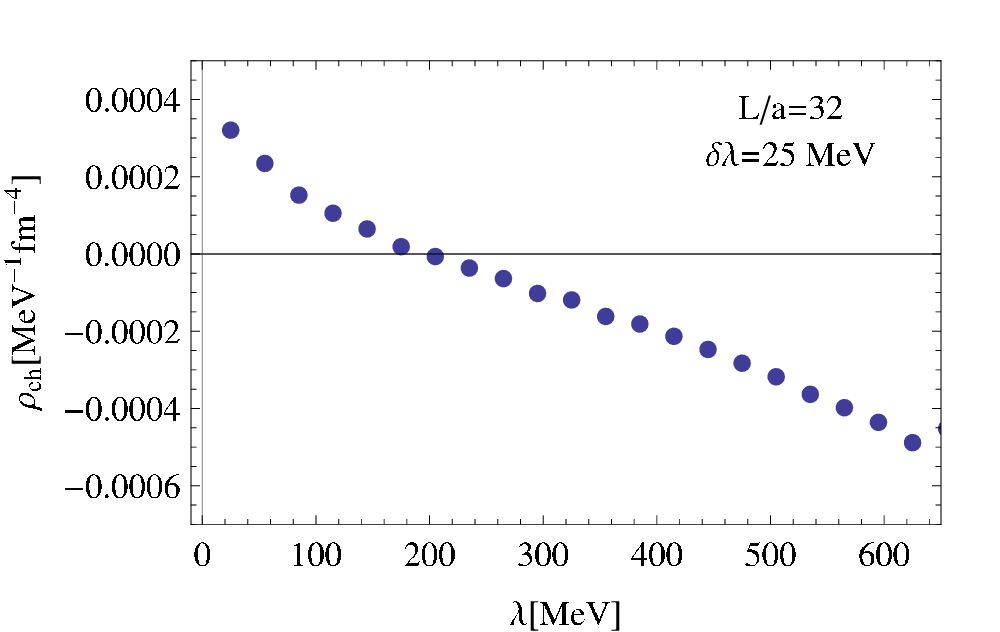}
     }
    \vskip -0.02in
    \centerline{
    \hskip 0.12in
    \includegraphics[width=0.52\textwidth,angle=0]{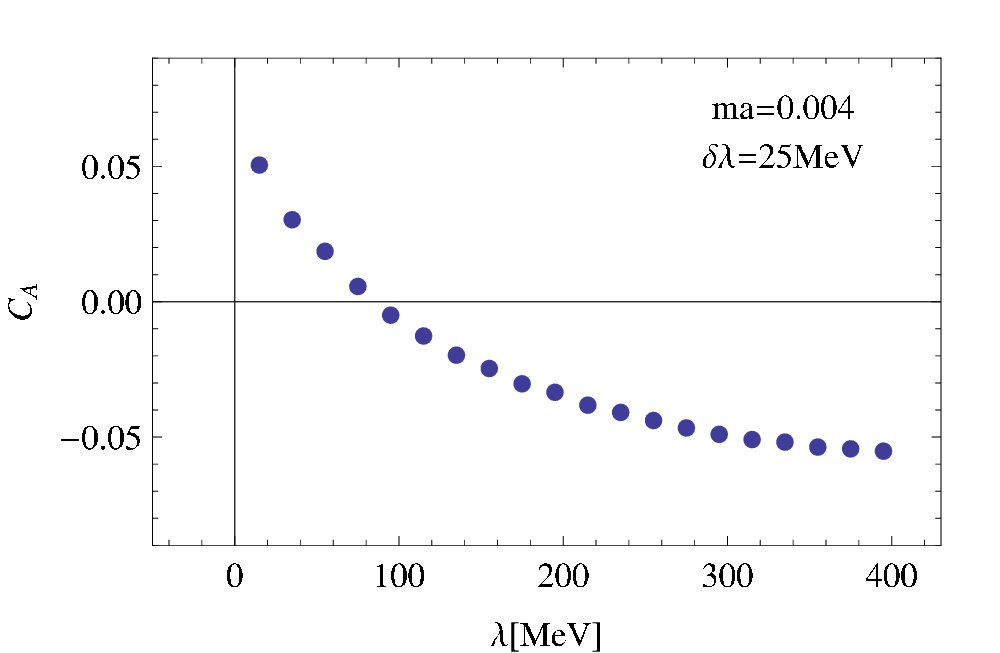} 
    \hskip -0.1in
    \includegraphics[width=0.52\textwidth,angle=0]{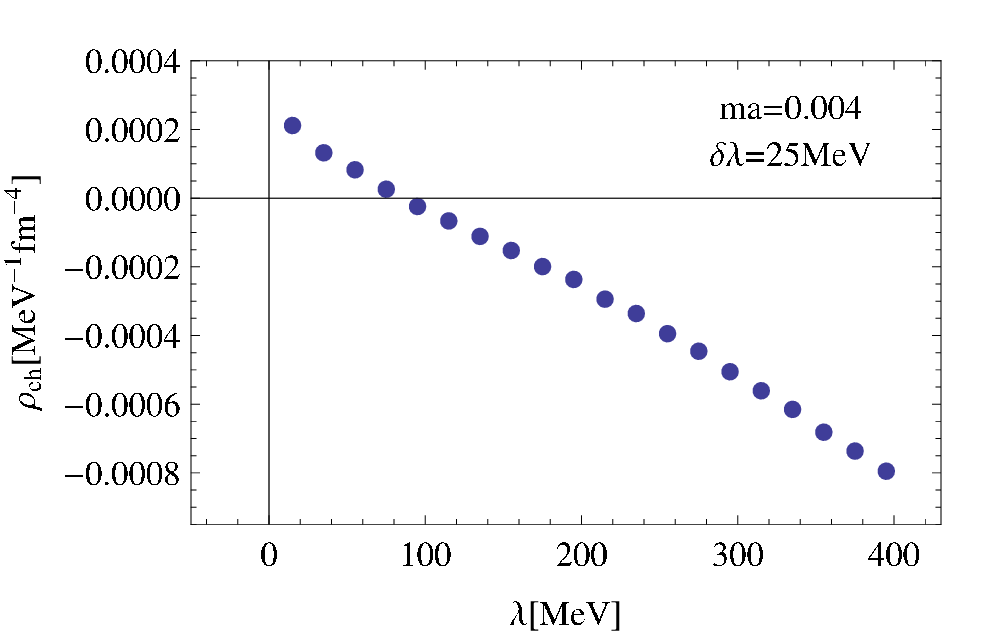}
     }
     \vskip 0.00in
     \caption{Top: behavior of $C_A(\lambda)$ and $\rho_{ch}(\lambda)$ in quenched QCD 
     (see text). Bottom: the same in N$_f$=2+1 QCD with domain wall fermions 
     and overlap Dirac probe (see text).} 
    \label{fig:2}
    \vskip -0.40in
\end{center}
\end{figure}

\section{The Reality of Chiral Polarization Scale}

The band of chirally polarized modes with associated $\Lambda_{ch}$ have first been 
seen in quenched QCD~\cite{Ale10A}, in a calculation at fixed physical volume. There is 
little doubt that quenched QCD is a mode--condensing theory, being studied in that regard 
since the early years of numerical lattice QCD (see e.g. Ref.~\cite{Bar84A}). However,
to establish the existence of local chirality condensation, and the reality of chiral 
polarization scale especially, an infinite volume asymptotics has to be examined. 

To do this, we have computed low--lying overlap--Dirac spectra on $16^4,20^4,24^4$ 
and $32^4$ lattices of quenched QCD with Wilson gauge action at $\beta=6.054$. 
Invoking reference scale $r_0=0.5\!$ fm, this corresponds to lattice spacing 
$a=0.085\!$~fm~\cite{Gua98A}. The parameters of the overlap Dirac operator  
in all spectral calculations discussed in this talk were set to $r=1$ and $\rho=26/19$. 
We computed 200 eigenmodes with smallest real part and  non--negative imaginary part 
of the eigenvalue, for 100 configurations from each ensemble. Correlation coefficient 
$C_A$ has then been calculated for each eigenmode. To evaluate the averages 
(\ref{eq:20}), (\ref{eq:30}) at given $\lambda$, the eigenmodes from the interval 
$(\lambda - \delta\lambda/2,\lambda + \delta\lambda/2)$ were used.

\begin{figure} 
  \centerline{
  \includegraphics[width=.7\textwidth]{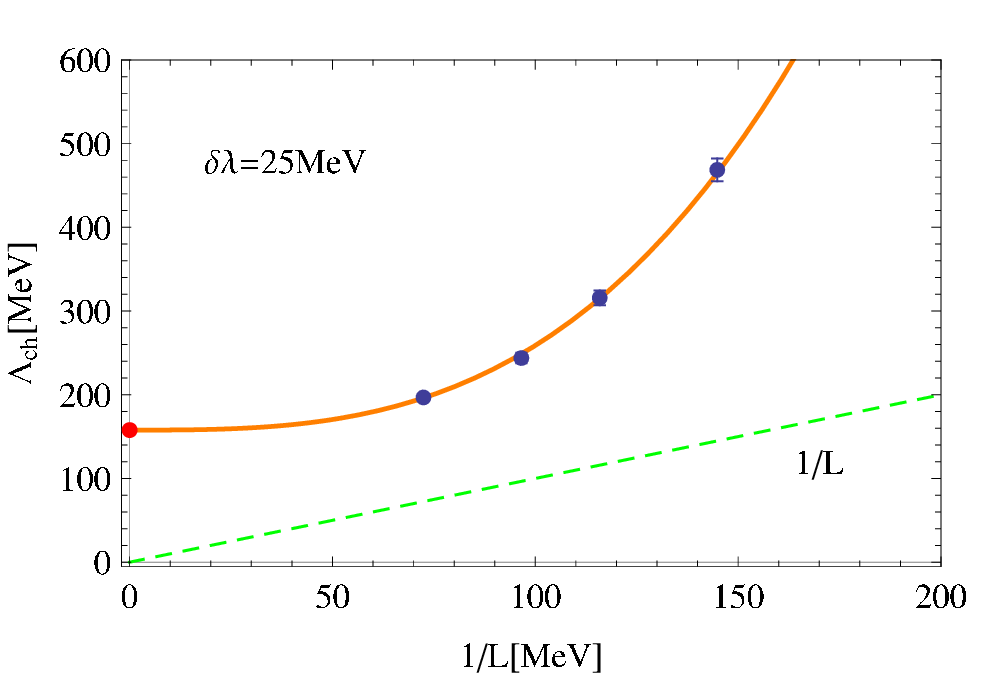}
  } 
  \caption{Volume dependence of $\Lambda_{ch}$ and its infinite volume extrapolation 
  in quenched QCD (see text).}
  \label{fig:3} 
  \vskip -0.10in
\end{figure}

In Fig.~\ref{fig:2} (top) we show the behavior of $C_A(\lambda)$ and $\rho_{ch}(\lambda)$
for the ensemble with largest volume. The scenario described in the Introduction 
is indeed observed with clearly defined band of chirally polarized modes and 
the associated $\Lambda_{ch}$. Note that since 
$ \rho(\lambda) = \rho_{ch}(\lambda) / C_A(\lambda) $ , the fact that both of the above
dependencies tend to non--zero value at small $\lambda$ is consistent with the expected
mode condensation property. Their positivity means that chirality condenses as well.
The volume dependence of $\Lambda_{ch}$ is shown in Fig.~\ref{fig:3}. The curvature
of the data away from the infrared cutoff, shown for comparison, suggests quite
convincingly that $\Lambda_{ch}$ is indeed a finite scale in the theory. The fit 
utilizing constant plus arbitrary power reveals a strong preference for cubic dependence,
and this power was then used to facilitate the infinite volume extrapolation shown.

\section{Light Dynamical Quarks}

Having established the viability of chiral polarization scale as a dynamically
generated feature of quenched QCD, it is now interesting to inquire whether
the proposed connection between mode condensation and chiral polarization of Dirac 
modes survives the effects of light dynamical quarks. The N$_f$=2+1 QCD at zero 
temperature is  
a suitable framework for such investigation both because it is close to the ``real world''
QCD when quark masses are adjusted accordingly, and because it is expected that this 
theory condenses for generic quark masses, including in the chiral limit.

To investigate this issue, we analyzed overlap Dirac eigenmodes in the $32^3\times 64$ 
ensembles of dynamical 
N$_f$=2+1 domain wall fermions generated by RBC/UKQCD collaborations~\cite{RBC09A}.
The quark mass parameters $M\equiv (m_l,m_l,m_h> m_l)$ in these ensembles have 
the heavy quark mass fixed at $m_h a =$ 0.03 while the light masses vary to be 
$m_l a =$ 0.008, 0.006, 0.004. The lattice scale ($a=$ 0.085 fm) was set via 
the physical value of the $\Omega$ baryon. Consequently, the heavy mass is fixed 
approximately at the strange quark value. The pseudoscalar meson masses 
associated with the three ensembles are $m_{\pi}=$ 397, 350 and 295 MeV.

\begin{figure} 
  \centerline{
  \includegraphics[width=.7\textwidth]{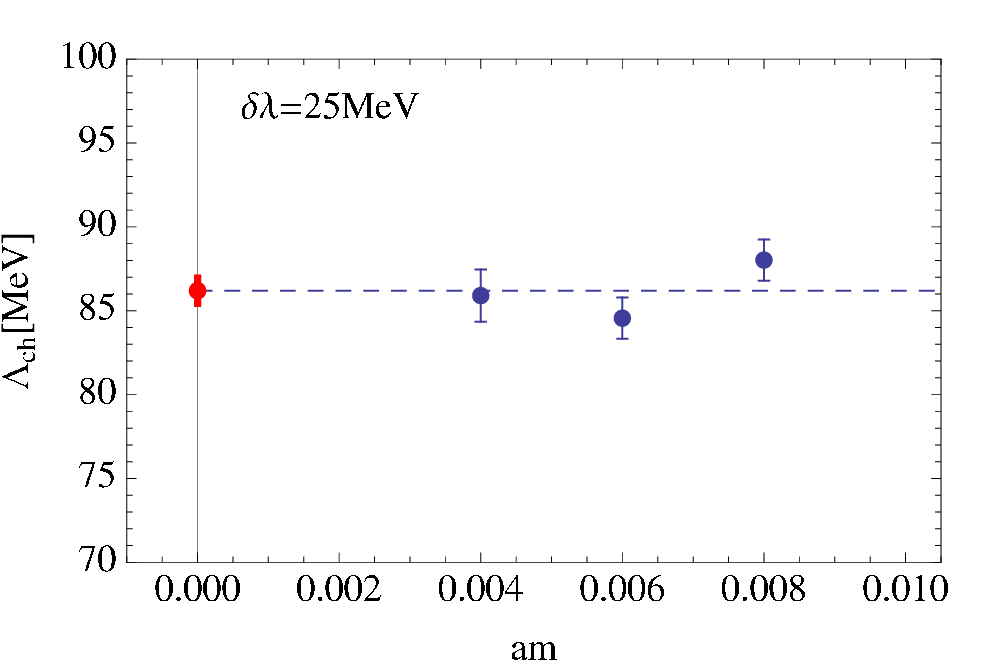}
  } 
  \caption{The light quark mass dependence of $\Lambda_{ch}$ in N$_f$=2+1 lattice QCD
  with domain--wall fermions and its chiral extrapolation via a constant (see text).}
  \label{fig:4} 
  \vskip -0.00in
\end{figure}

In Fig.~\ref{fig:2} (bottom) we plot the functions $C_A(\lambda)$ and $\rho(\lambda)$
for the ensemble with smallest light quark mass. We observe the behavior qualitatively
similar to the quenched case, albeit with chiral polarization scale that is somewhat
smaller. The situation for the heavier light quark masses is the same, even quantitatively.
Indeed, to assess the trend in the direction of chiral limit, we show 
in Fig.~\ref{fig:4} the mass dependence of $\Lambda_{ch}$ which turns out 
to be entirely flat. This suggests that chiral limit may not induce any dramatic effects 
beyond lowering $\Lambda_{ch}$ relative to quenched theory, as is already apparent 
in our data.

The experience with continuum limit extrapolation at fixed volume~\cite{Ale10A}, combined
with trends in infinite volume extrapolation shown here, suggest the value of chiral 
polarization scale $\Lambda_{ch}\approx$ 75--80 MeV both in the chiral limit and at 
the physical point. The existence of $\Lambda_{ch}$ is expected to be independent
of the regularization (lattice action) used to define the continuum N$_f$=2+1 theory, 
but it remains to be seen how large the regularization--dependent spread in these 
values is. Nevertheless, given that $\Lambda_{ch}$ is a dynamical characteristic, 
the variation may in fact be very small.

\section{Discussion}

The main message of this work is that quark--gluon setups relevant for 
``real world'' at zero temperature, such as N$_f$=2+1 QCD, have Dirac spectra with
striking dynamical feature: there is a band of chirally polarized low--energy modes 
extending up to a well--defined dynamical scale $\Lambda_{ch}$. 
Such behavior appears to take place at generic quark masses, including in the chiral 
limit, thus being a feature of spontaneous chiral symmetry breaking in that theory. 
This observation provides a useful constraint for possible model descriptions of low 
energy QCD, but also a necessary ingredient
in deeper understanding of SChSB's dynamical origin. Indeed, since quark dynamics 
at small light masses is dominated by low--lying modes
we can conclude that chirally broken quark dynamics is ``facilitated'' by chirally
polarized modes.

As can be seen from our discussion, chiral polarization of modes does not arise due 
to light quarks. In fact, light quarks make it milder. In that regard, chiral polarization
appears to be tied to mode condensation property (valence chiral condensate) 
which is strongest in pure glue QCD and quarks effects weaken it since they slow down 
the running of coupling toward larger values at low energy. 
As emphasized in Ref.~\cite{Ale12D}, this meshes well with presumed free--like behavior 
of modes at high energies, since free fermions are perfectly 
chirally anti--polarized. Turning on the gauge interaction weakens this order even 
at high energies, but potentially reverses the dynamical trend at long distances 
due to rising coupling. The observed existence of chirally polarized band and 
$\Lambda_{ch}$ in quenched QCD means that this indeed happens: chiral dynamical properties
of modes qualitatively change at $\Lambda_{ch}$, with modes becoming chirally polarized
and near--zeromodes condensing.
Light quark effects in ``real world'' QCD at zero temperature are not able to reverse this, 
resulting in SChSB via condensation of dynamical chirality. 

Given the above considerations, it is hard to avoid thinking of $\Lambda_{ch}$ as a natural 
scale of broken chiral dynamics, be it current quarks or valence: it is a scale at which 
the representative of quark dynamics, namely Dirac eigenmode at that scale, starts 
chirally resembling a representative of purely broken quark dynamics, namely a
strict near--zeromode. This view would be further strengthened if mode condensation and 
chiral polarization only occurred simultaneously within the realm of QCD--like
theories, which we take to be SU(3) gauge theories with arbitrary number of
fermionic species in fundamental representation, and at arbitrary temperature. Indeed, 
if this were true then the proposed role of chiral polarization would become inherent 
to quark--gluon interaction itself, with $\Lambda_{ch}$ acquiring a fundamental 
meaning directly related to mechanism of SChSB. In Ref.~\cite{Ale12D} we have conjectured 
that the above mode condensation--chiral polarization equivalence indeed holds. While there 
are many corners for explicit checks of this relationship, it appears to hold with regard 
to thermal agitation~\cite{Ale12D,Ale12E,Ale12F}, at least in quenched QCD.

The above scenario has interesting conceptual and practical implications discussed in some 
detail in Ref.~\cite{Ale12D}. They mostly have to do with novel ways of characterizing
broken chiral dynamics. Here we limit ourselves to pointing
out that the proposed dynamical insight turns chiral polarization scale 
$\Lambda_{ch}$ into a non--traditional ``order parameter'' of SChSB. Among other
things, this has a practical utility in that it is a strictly well--defined concept 
even at finite volume. Thus, for a given finite system, one can uniquely (and cheaply) 
delineate the regions of parameter space characterized by chiral polarization and 
non--zero $\Lambda_{ch}$. These then turn into the regions of broken chiral dynamics,
or regions of mode condensation, in the infinite volume limit.


\begin{thebibliography}{99}

\bibitem{Ale12D} A.~Alexandru, I.~Horv\'ath, {\tt arXiv:1210.7849}.

\bibitem{Ban80A} T.~Banks and A.~Casher, Nucl.~Phys. {\bf B169}, 125 (1980).

\bibitem{Hor01A} 
   I.~Horv\'ath, N.~Isgur, J.~McCune, H.B.~Thacker, Phys.~Rev.~{\bf D65}, 014502 (2002),
   {\tt hep-lat/0102003}.

\bibitem{Neu98BA}
   H.~Neuberger, Phys.~Lett. {\bf B417} (1998) 141; 
                 Phys.~Lett. {\bf B427} (1998) 353.

\bibitem{Ale10A} 
   A.~Alexandru, T.~Draper, I.~Horv\'ath, T.~Streuer, 
   Annals Phys. {\bf 326}, 1941 (2011), {\tt arXiv:1009.4451}.

\bibitem{Hor06A} I.~Horv\'ath, {\tt hep-lat/0605008}.

\bibitem{Bar84A}
   I.~Barbour {\em et al.}, Phys.~Lett. {\bf B136} (1984) 80. 

\bibitem{Gua98A}
   {\bf ALPHA} Collaboration, M.~Guagnelli, R.~Sommer and H.~Wittig, 
   Nucl.~Phys. {\bf B535}, 389 (1998), {\tt hep-lat/9806005}.

\bibitem{RBC09A} R.~Mawhinney, ({\bf RBC} and {\bf UKQCD} Collaborations) 
  {\sl PoS} {\bf LAT2009}, 81 (2009), {\tt arXiv:0910.3194} 

\bibitem{Ale12E} A.~Alexandru, I.~Horv\'ath,
  $\,$ PoS (LATTICE 2012) (2012) 210, {\tt arXiv:1211.2601}.

\bibitem{Ale12F} A.~Alexandru, I.~Horv\'ath,
  $\,$ Proceedings of the international workshop, ``Extreme QCD'', Washington, DC,
  Aug 21-23, 2012. Submitted. {\tt arXiv:1211.3728}.


\end{thebibliography}
\end{document}